\documentclass[prl,twocolumn,%
               preprintnumbers,
               nofootinbib,floatfix,
               showpacs]{revtex4}    % LaTeX 2e
\usepackage{amsmath}
\usepackage{amsfonts}
\usepackage{graphicx}
\usepackage{psfrag}

\newcommand{\vect}[1]{\vec{#1}}
\newcommand{\op}[1]{\hat{#1}}
\renewcommand{\d}{\mathrm{d}}

\newcommand{\pdiff}[3][{}]{\frac{\partial^{#1}{#2}}{\partial{#3}^{#1}}}
\newcommand{\abs}[1]{\lvert{#1}\rvert}
\newcommand{\braket}[1]{\langle{#1}\rangle}
\newcommand{\ket}[1]{|{#1}\rangle}

\begin{document}

\date{\today}
\title{Stability Criteria for Breached Pair Superfluidity}
\author{Michael McNeil Forbes, Elena Gubankova, W. Vincent Liu, Frank Wilczek}
\affiliation{Center for Theoretical Physics, Department of Physics,
  MIT, Cambridge, Massachusetts 02139}
\preprint{MIT-CTP-3491}

\begin{abstract}
  We present simple, concrete, two-fermion models that exhibit
  thermodynamically stable isotropic translationally-invariant gapless
  superfluid states (breached pair superfluidity).  The mass ratio
  between the components and the momentum structure of the interaction
  are crucial for determining the stability of such states: Idealized,
  momentum-independent (``contact'') interactions are insufficient.
\end{abstract}

\maketitle
\section{Introduction}\noindent
Recently there has been interest in superfluid fermion systems where
there exist superfluid states that retain gapless fermionic
excitations~\cite{Liu:2002gi,Gubankova:2003uj,Shovkovy:2003uu,Alford:2003fq}.
These states embody ``phase separation in momentum space'': some
degrees of freedom pair, forming a superfluid, while others remain
unpaired, maintaining the properties of a Fermi surface.  They are
likely to become experimentally accessible in the near
future~\cite{Liu:2004mh}.

The important new result is that such states are stable, but to see
this, one must perform a careful analysis of the momentum dependence
of the interaction.  Appropriate systems (such as two-band
models~\cite{SMW:1959,Kondo:1963}) have been studied, but these states were
missed because the possibility of momentum dependence was ignored.  We
also address the questions of instability toward phase
separation~\cite{Bedaque:2003hi} and local currents~\cite{Wu:2003}
raised in response to the proposal~\cite{Liu:2002gi}.

We shall consider systems of two species that, in the absence of
interactions, would have two distinct Fermi surfaces.  Simple
heuristic considerations suggest the possibility that pairing takes
place about the Fermi surfaces, but that there is no pairing in a
region between the surfaces: this led to the term ``breached
pair''~\cite{Gubankova:2003uj}.  A breached pair superfluid state (BP)
is characterized by the coexistence of a superfluid and a normal
component in a translationally invariant and isotropic state.  These
components are accommodated in different regions of momentum space
with the normal component residing in the ``breaches'', bounded by
gapless Fermi surfaces.

A state of this type was also considered by Sarma~\cite{Sarma:1963}.
He considered the case of a superconductor in an external magnetic
field, and found that, although there is a self-consistent mean-field
solution with gapless modes, it is unfavored energetically to the
fully gapped BCS solution.  Similar results were considered in the
context of color superconductivity~\cite{Alford:1999xc}, again
concluding that these states are not stable at fixed chemical
potential.

Since the fully gapped BCS solution enforces equal numbers of each
species, it was incorrectly suggested~\cite{Gubankova:2003uj} that one
might stabilize the gapless phase by enforcing constraints on the
particle numbers.  Indeed, by enforcing unequal numbers of each
species, one forbids the formation of a fully gapped BCS state, but
admits ``breached pair'' states in which the excess in one species can
be accommodated by the breach.  In the QCD context, a similar argument
has been made by imposing charge
neutrality~\cite{Shovkovy:2003uu,Alford:2003fq}.  Recently, however,
Bedaque, Caldas and Rupak~\cite{Bedaque:2003hi,Caldas:2003ig} pointed
out that a spatially mixed phase may be energetically preferable: this
rules out the first possibility~\cite{Gubankova:2003uj} but may not
affect the QCD case due to the long-range gauge interactions.

Here we clarify, broaden, and correct this discussion.  We conclude
that:
\begin{itemize}
\item For extensive systems, one can not stabilize a state by imposing
  different global constraints (such as fixed particle number): The
  composition of the state can be completely determined from an
  analysis of the grand canonical ensemble.  The specific examples
  considered in~\cite{Gubankova:2003uj} are accordingly unstable.
\item With the proper momentum structure, however, one may realize
  breached pair superfluidity in states that are thermodynamically
  stable for fixed chemical potentials.  We exhibit these below.
\end{itemize}
Our considerations do not apply directly to non-extensive systems.
Charge conservation or color neutrality constraints enforced by
long-range gauge forces might stabilize BP phases.  (Of course, the
possibility of a competing mixed phase must still be considered
quantitatively.)

\section{Thermodynamic Stability}\noindent
In the context of two component fermionic systems as considered
in~\cite{Gubankova:2003uj,Bedaque:2003hi,Caldas:2003ig},
three competing homogeneous phases have been considered: a normal
state of free fermions (N), a fully gapped superfluid phase (BCS), and
a gapless BP phase.  The BCS phase has complete pairing between the
two species, and thus enforces equal densities.  The other phases
admit differing densities.

Upon solving the self-consistency conditions (gap equations), one
commonly finds that over a range of chemical potentials there are
three distinct solutions.  To determine which is stable in this grand
canonical ensemble, one must minimize the grand thermodynamic
potential (equivalently, maximize the pressure) of the system.
Typically, two of the three solutions are minima on either side of the
third BP state which is a local maximum: Fig.~\ref{fig:Dispersion}
shows a typical potential.  The gapless states found
in~\cite{Sarma:1963,Gubankova:2003uj} correspond to local maxima, thus
the competing state with larger gap parameter $\Delta$ has higher
pressure and renders the BP state unstable in this ensemble.

If the stable solution is fully gapped, then it has equal densities.
By fixing the particle densities to be unequal, one may forbid this
BCS state.  Furthermore, upon comparing the Helmholtz free energies
$H$---which must be minimized in this ensemble---one may find that the
``unstable'' BP state is favored over the normal state N.

\begin{figure}[ht]
  \begin{centering}
    \psfrag{Omega}{$\frac{\Omega(\mu)}{V}$}
    \psfrag{mu0}{$\mu_0$}
    \includegraphics[width = 0.3\textwidth]{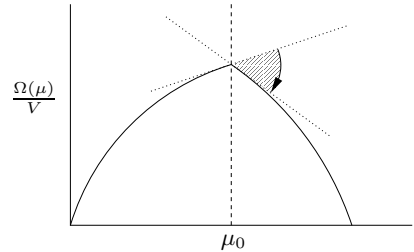}
  \end{centering}
  \caption{\label{fig:tangents}
    The cone of tangent (hyper)planes to a thermodynamic potential
    density $-P = \Omega(\mu)/V$.  Immediately to the left of
    $\mu_0$ is a pure phase with density $n_L$ while immediately to
    the right is another pure phase with density $n_R$.  The
    densities are the negative slopes of the tangents at $\mu_0$
    according to~(\ref{eq:N}).  At $\mu=\mu_0$ there is a continuum
    of mixed phases: These consist of a volume fraction $x$ at
    density $n_L$ and the remaining fraction $1-x$ at density $n_R$.
    The average density over all space, $n = x n_L+(1-x)n_R$, lies
    within $n\in (n_L,n_R)$.}
\end{figure}

This apparent contradiction in the stability analysis based on
different ensembles can be resolved by considering a mixed
phase~\cite{Bedaque:2003hi,Caldas:2003ig} which has an even lower
Helmholtz free energy $H$.  That such a resolution is always
possible, however, may not be apparent; indeed, it is generally hard
to determine the mixed phase explicitly.  By using general
properties~\cite{Sewell:2002} of the grand thermodynamic potential
$\Omega$, however, one can argue that such a solution is always possible,
as follows.  By definition,
\begin{equation}
  \label{eq:OmegaDef}
  -PV = \Omega(\vect{\mu}) = \min(H-\vect{\mu}\cdot\vect{N}),
\end{equation}
minimized over all competing phases.  Thus, $\Omega$ is a concave
function of the chemical potentials $\vect{\mu} = (\mu_a,\mu_b)$.  (We
consider here fixed $T=0$, but concavity in $T$ also follows from
maximizing entropy.)  Furthermore, there is a one-to-one
correspondence between tangents $\Omega$ and states of fixed particle
number:
\begin{equation}
  \label{eq:N}
  \vect{N} = -\pdiff{\Omega}{\vect{\mu}}.
\end{equation}
When $\Omega$ is not differentiable, there is a cone of possible
tangent hyperplanes which contact $\Omega$ and which bound $\Omega$
from above (see Fig.~\ref{fig:tangents}).  This cone of tangents
describes various possible mixed phases composed of the pure phases
(where $\Omega$ is differentiable) that intersect at the singularity.
To find the state that minimizes $H$ for some fixed constraint
$\vect{N}=\vect{N}_0$ one simply forms the hyperplane with gradient
$\vect{N}_0$ and drops this until it contacts the surface $\Omega$.
The first point of contact will define either a pure or mixed state
which satisfies the appropriate constraints.  Since this state also
lies on $\Omega$, it minimizes $\Omega$ for the fixed chemical
potentials defined by the contact point.  No matter what constraints
we apply, there is always a stable state in the grand canonical
ensemble.

This argument is valid only for extensive thermodynamic systems.
Long-range interactions can render the energy of some pure phases
non-extensive (due, for example, to the rapidly diverging Coulomb
energy per unit volume $V$ as $V \rightarrow \infty$).  In such cases,
a mixed phase would contain bubbles of limited size.  The surface
energy of these phase boundaries becomes a volume effect and must
therefore be taken into account, even in the thermodynamic limit (see
for example~\cite{Reddy:2004my}).  This complicates the relation
between $\vect{N}$ and $\vect{\mu}$.  In this rest of this letter, we
shall consider only finite-range interactions.  
\section{Stable Breached Pair Superfluids}\noindent 
We now demonstrate, by example, how to realize pure BP superfluid
states in extensive systems.  We shall consider the mean-field analysis
of two models, each with two species of fermions $a$ and $b$ of
differing masses $m_a < m_b$:
\begin{equation}
  \label{eq:H}
  \mathcal{H} = \int\frac{\d^3{\vect{p}}}{(2\pi)^3}
  \left(
    \frac{p^2}{2m_a}
    \op{a}^\dagger_{\vect{p}}
    \op{a}_{\vect{p}} + 
    \frac{p^2}{2m_b}
    \op{b}^\dagger_{\vect{p}}
    \op{b}_{\vect{p}}
  \right)
  + \mathcal{H}_I.
\end{equation}
The models have different interactions $\mathcal{H}_I$: (\ref{eq:Vr})
and~(\ref{eq:V}).  We shall consider these systems in the grand
canonical ensemble at zero temperature by minimizing the thermodynamic
potential density $\Omega(\mu_a,\mu_b)/V$.  It will be natural,
however, to use the parameters $p^F_{i}=\sqrt{2m_{i}\mu_{i}}$ in
place of the chemical potentials $\mu_i$.

The first model posits a spherically symmetric static two-body
potential interaction $V(r)$ between the two species $a$ and $b$:
\begin{equation}
  \label{eq:Vr}
  \mathcal{H}_{I} = 
  \int \d^{3}{\vect{x}}\;\d^{3}{\vect{y}}\;\;
  V\bigl(\abs{\vect{x}-\vect{y}}\bigr)\;\;
  \smash{\op{\vphantom{b}a}}^\dagger_{\vect{x}}\;
  \op{b}^\dagger_{\vect{y}}\;
  \op{b}_{\vect{y}}\;
  \op{\vphantom{b}a}_{\vect{x}}.
\end{equation}
Defining $m_\pm = 2m_am_b/(m_b\pm m_a)$, $\mu_\pm =
(\mu_a\pm\mu_b)/2$, and
\begin{equation}
  \epsilon^{\pm}_{p} \equiv \tfrac{1}{2}
  \left[
      \frac{p^2}{2m_a} - \mu_a\right]
    \pm
    \tfrac{1}{2}
    \left[
      \frac{p^2}{2m_b} - \mu_b\right]
    = \frac{p^2}{2m_\pm} -\mu_{\pm},
\end{equation}
and considering only homogeneous (translationally invariant) and
isotropic phases, we find that extrema of~(\ref{eq:OmegaDef}) satisfy
the gap equation
\begin{equation}
  \label{eq:GapEqP}
  \Delta_{p} = 
  -\int_R \frac{\d^{3}\vect{q}}{(2\pi)^3}\;\;
  \tilde{V}\bigl(\abs{\vect{p}-\vect{q}}\bigr)\frac{\Delta_q}
  {2\sqrt{(\epsilon^+_q)^2+\Delta_q^2}},
\end{equation}
where $\tilde{V}(p)$ is the Fourier transform of $V(r)$.  The
integral~(\ref{eq:GapEqP}) runs over the region $R$ outside any
``breach''.  $R$ contains momenta where the two quasiparticle
dispersions $E^{\pm}_p$
\begin{equation}
  \label{eq:LambdaP}
  E^{\pm}_p = \epsilon^{-}_{p} \pm
  \sqrt{(\epsilon^+_p)^2+\Delta^2_{p}},
\end{equation}
have opposite sign.  (See~\cite{Gubankova:2003uj} for further details
about the generic breach structure.)  Note from~(\ref{eq:GapEqP}) that
$\Delta_p$ is generally largest about $p_0$, where $\epsilon^{+}_{p_0}
= 0$.

Equation~(\ref{eq:GapEqP}) can be solved numerically to find extremal
points of the thermodynamic potential.  Over this set of
self-consistent solutions, one can minimize $\Omega$ to determine the
phase structure.

\begin{figure}[t]
  \begin{centering}
    \psfrag{5}{$5$}
    \psfrag{10}{$10$}
    \psfrag{15}{$15$}
    \psfrag{20}{$20$}
    \psfrag{pa}{$p^F_a$}
    \psfrag{pb}{$p^F_b$}
    \psfrag{BCS}{BCS}
    \psfrag{BP}{BP}
    \psfrag{Normal}{Normal}
    \includegraphics[width=8.6cm]{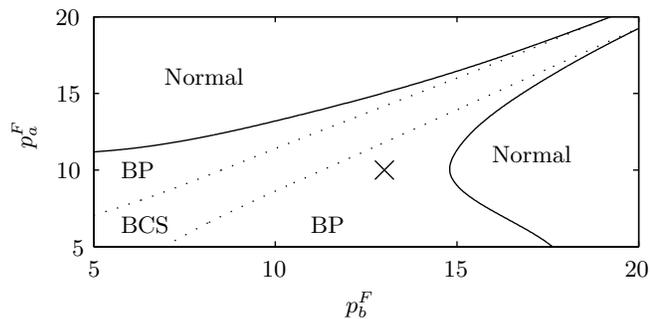}
    \caption{\label{fig:phaseP}
      Qualitative $T=0$ phase diagram for interaction~(\ref{eq:Vr})
      with a Gaussian potential $V(r) \propto \exp(-r^2/2\lambda^2)$.
      All momentum scales are in units of $\hbar/\lambda$ and all
      energy scales are in units of $\hbar^2/(m_+\lambda^2)$.  The
      mass ratio is $m_b/m_a = 50$ and the coupling strength has been
      chosen so that $2m_+\Delta_{p_0}/p_0^2 = 0.1$ at the point
      marked ``$\times$'' where $(p^F_b,p^F_a) = (13,10)$ to ensure
      weak-coupling.  (This ratio is less that 1 throughout this
      diagram.)  Note that the lower BP region has more heavy
      particles $b$ while the upper BP region has fewer heavy
      particles.  The upper type may be realized in the QCD
      context~\cite{Alford:2003fq}.  All phase transitions are first
      order.}
  \end{centering}
\end{figure}

\begin{figure}[t]
  \begin{centering}
    \psfrag{-6}{$-6$}
    \psfrag{0}{$0$}
    \psfrag{+1}{$\hphantom{+}1$}
    \psfrag{+0}{$\hphantom{+}0$}
    \psfrag{+2}{$\hphantom{+}2$}
    \psfrag{+4}{$\hphantom{+}4$}
    \psfrag{6}{$6$}
    \psfrag{8}{$8$}
    \psfrag{10}{$10$}
    \psfrag{12}{$12$}
    \psfrag{14}{$14$}
    \psfrag{l}{$E^{\pm}_{p}$}
    \psfrag{na,nb}{$n_a$, $n_b$}
    \psfrag{na=nb}{$n_a=n_b$}
    \psfrag{na}{$n_a$}
    \psfrag{nb}{$n_b$}
    \psfrag{Dp}{$\Delta_p$}
    \psfrag{p}{$p$}
    \includegraphics[width=8.6cm]{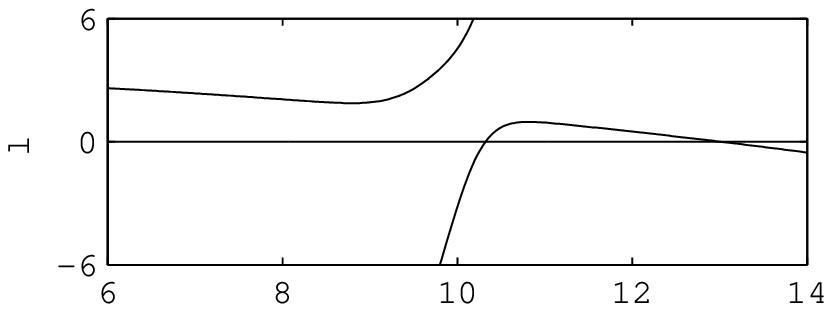}\\
    \includegraphics[width=8.6cm]{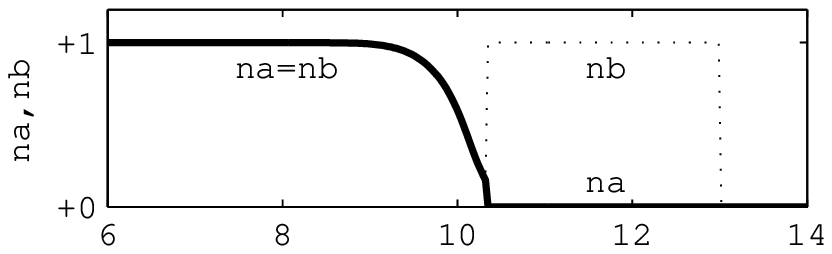}\\
    \includegraphics[width=8.6cm]{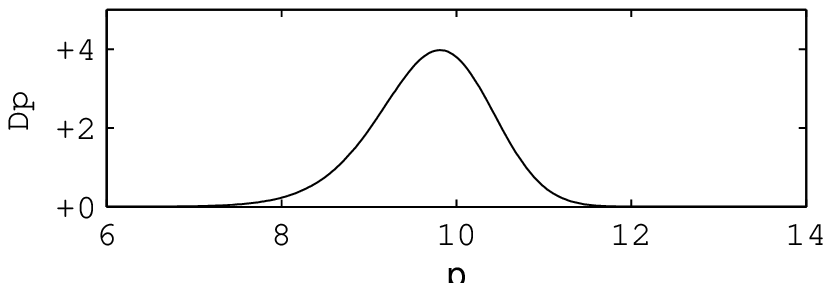}
    \caption{\label{fig:DispersionP} Quasi-particle dispersions
      $E^\pm_{p}$ (top), occupation numbers $n_a$ and $n_b$ (middle),
      and gap parameter $\Delta_{p}$ for a sample BP state at
      $(p^F_b,p^F_a) = (13,10)$.  All units and parameters are
      described in Fig.~\ref{fig:phaseP}.  Notice that there are two
      ``Fermi'' surfaces at $p\approx 10.3$ and $p=13$.  The first
      occurs where $\Delta_{p}$ becomes too small to support the gap.
      The second is simply the Fermi surface for $b$ which is
      virtually unaffected by the pairing.  The ``breach'' occurs
      between these surfaces and only the region $R$ outside
      contributes to the gap equation~(\ref{eq:GapEqP}).}
  \end{centering}
\end{figure}

We have done this for a variety of interactions, and find similar
qualitative structure: a central strip of fully gapped BCS-like phase
about $p^F_a=p^F_b$, with normal unpaired phases outside (see
Fig.~\ref{fig:phaseP}.)  Depending on the model parameters, these
phases may be separated by a region of BP superfluid phase.  To verify
that these indeed contain gapless modes we plot in
Fig.~\ref{fig:DispersionP} a sample set of occupation numbers,
quasiparticle dispersions, and the gap parameter $\Delta_p$,
\begin{equation}
    \Delta_p = \int\frac{\d^3{\vect{q}}}{(2\pi)^3} V(\abs{\vect{p}-\vect{q}})
    \braket{\op{b}_{\vect{p}}\op{a}_{-\vect{p}}}.
  \end{equation}

The presence of gapless fermion modes depends crucially on two
factors: 1) the momentum structure of $\Delta_{p}$ and, 2) the mass
ratio.

First, $\Delta_p$ must be large in some regions and small in others.
If $\Delta_p$ is large enough at a Fermi surface, it will induce
pairing at that surface and support a superfluid.  If it is also small
enough at the other Fermi surface, it will not appreciably affect the
normal free-fermion behaviour.  The problem with previous analyses is
that they assume pointlike interactions, implying a constant
$\Delta_{p} = \Delta$.  Physical interactions, however, tend to
exhibit more complicated behaviour and suitable $\Delta_{p}$ are quite
generic: $\Delta_{p}$ tends to peak about the Fermi surface of the
lighter species and fall off to at least one side.  The model shown in
Fig.~\ref{fig:phaseP} and Fig.~\ref{fig:DispersionP}, for example, has
a Gaussian interaction.  Longer-range forces (such as a screened
Coulomb interaction) tend to plateau to the left of $p_0$ but still
fall to the right of $p_0$.

Second, as was emphasized in~\cite{Liu:2002gi}, the one may reduce the
cost associated with shifting the Fermi sea $p^F_b$ by increasing the
mass $m_b$.  Thus, by choosing a large enough mass ratio, one may
always move the Fermi surface for the heavy species to region where
$\Delta_{p^F_b}$ is small enough to leave the Fermi surface
undisturbed. The states shown in Fig.~\ref{fig:phaseP} and
Fig.~\ref{fig:DispersionP} have a mass ratio $m_b/m_a = 50$.

Since the variational states of model~(\ref{eq:Vr}) are parameterized
by a variable \emph{function} $\Delta_{p}$, the set of states over
which the minimization~(\ref{eq:OmegaDef}) must consider is
enormous, and we cannot be certain to have found the global minimum.
We have searched for stable fixed-points of the gap
equation~(\ref{eq:GapEqP}) and compared them, so our results for this
model are consistent and plausible, but not rigorous.

The second model allows us to be rigorous.  We fall back to the type
of factorized, cutoff interaction often considered in BCS models:
\begin{equation}
  \label{eq:V}
  \mathcal{H}_{I} = 
  g \int \frac{\d^{3}{\vect{p}}}{(2\pi)^3}
  \frac{\d^{3}{\vect{q}}}{(2\pi)^3}\;
  f(p)f(q)\;
  \smash{\op{\vphantom{b}a}}^\dagger_{\vect{p}}\;
  \op{b}^\dagger_{-\vect{p}}\;
  \op{b}_{-\vect{q}}\;
  \op{\vphantom{b}a}_{\vect{q}}.
\end{equation}
In this model, the combination $\Delta f(p)$ plays the same role as
$\Delta_p$ in first model, with the cutoff function $f(p)$ providing the
required momentum dependence.  The trial states are now parameterized
by a single number $\Delta = \braket{\op{\Delta}}$ where $\op{\Delta}
= g\int\d^{3}{\vect{p}}/(2\pi)^3 f(p)\op{b}_p\op{\vphantom{b}a}_{-p}$.
\begin{figure}[t]
  \begin{centering}
    \psfrag{4}{$4$}
    \psfrag{6}{$6$}
    \psfrag{8}{$8$}
    \psfrag{10}{$10$}
    \psfrag{12}{$12$}
    \psfrag{BCS}{BCS}
    \psfrag{BP}{BP}
    \psfrag{Normal}{Normal}
    \psfrag{pa}{$p^F_a$}
    \psfrag{pb}{$p^F_b$}
    \includegraphics[width=8.6cm]{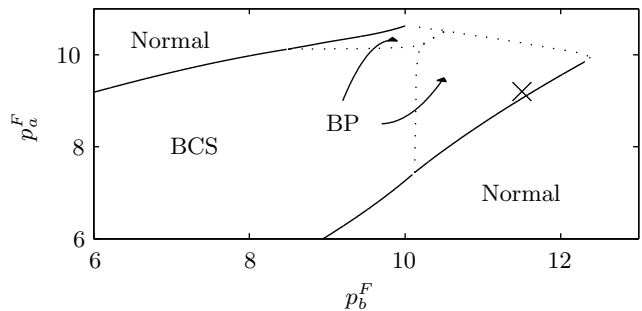}
    \caption{\label{fig:phase} $T=0$ phase diagram for
      model~(\ref{eq:V}) with a hard cutoff $f(p) \sim
      \theta(p-10\sigma)$ that has been smoothed over the range from
      $p \in (9.7,10.3)$.  All momenta are expressed in units of
      $\sigma$ where $10\sigma$ is the cutoff scale, and all energies
      are expressed in units of $\sigma^2/(2m_+)$.  The mass ratio is
      $m_b/m_a=4$ and the coupling $g$ has been chosen so that
      $2m_+\Delta/p_0^2 = 0.2$ at $p^F_a = p^F_b = p_0 = 10\sigma$ to
      ensure weak-coupling.  (This ratio is less that $1$ throughout
      this diagram).  All phase transitions are first order as
      discussed in the text. The mixed phases
      of~\cite{Bedaque:2003hi,Caldas:2003ig} would be found on the
      solid lines.  The sample state in Fig.~\ref{fig:Dispersion} at
      $(p^F_b,p^F_a) = (11.5,9.2)$ is marked ``$\times$''.}
  \end{centering}
\end{figure}
\begin{figure}[t]
  \begin{centering}
    \psfrag{-20}{$-20$}
    \psfrag{-2}{$-2$}
    \psfrag{-1}{$-1$}
    \psfrag{-00}{$\hphantom{-0}0$}
    \psfrag{+01}{$\hphantom{+0}1$}
    \psfrag{0}{$0$}
    \psfrag{1}{$1$}
    \psfrag{2}{$2$}
    \psfrag{5}{$5$}
    \psfrag{10}{$10$}
    \psfrag{15}{$15$}
    \psfrag{20}{$20$}
    \psfrag{40}{$40$}
    \psfrag{1}{$1$}
    \psfrag{7}{$7$}
    \psfrag{13}{$13$}
    \psfrag{p}{$p$}
    \psfrag{lambda}{$E^{\pm}_p$}
    \psfrag{na,nb}{$n_a$, $n_b$}
    \psfrag{na=nb}{$n_a=n_b$}
    \psfrag{na}{$n_a$}
    \psfrag{nb}{$n_b$}
    \psfrag{Delta}{$\Delta$}
    \psfrag{FreeEnergyDif}{$\Bigl(\Omega(\Delta)-\Omega(0)\Bigr)/V$}
    \includegraphics[width=8.6cm]{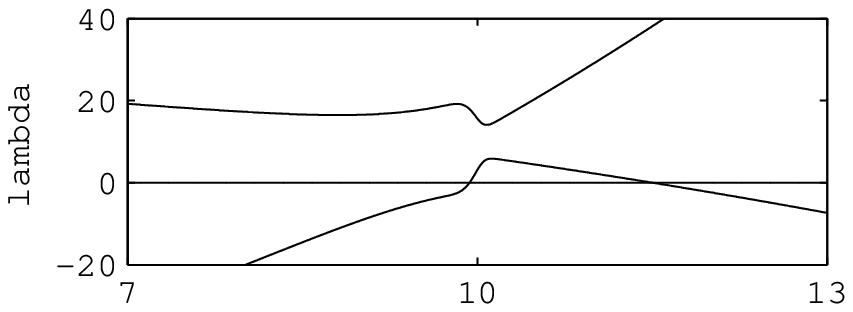}
    \includegraphics[width=8.6cm]{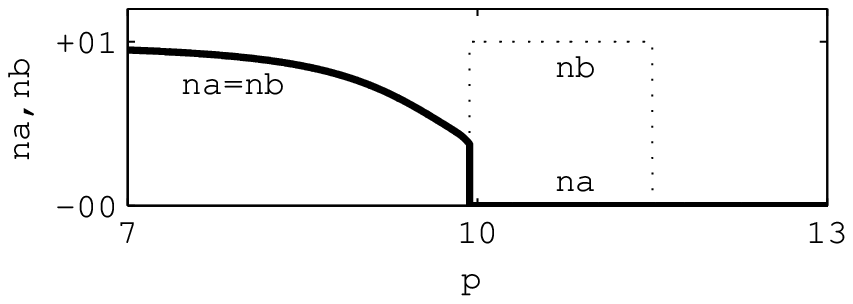}
    \includegraphics[width=8.6cm]{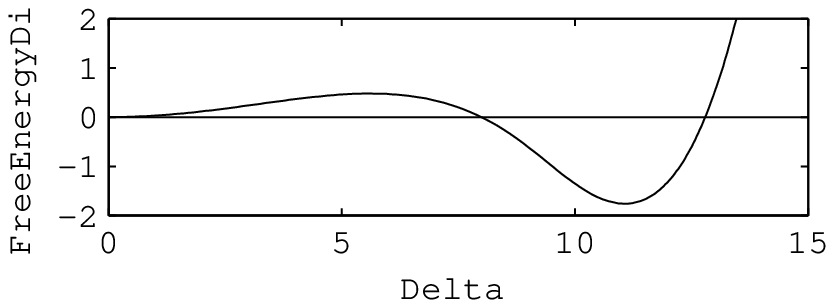}
    \caption{\label{fig:Dispersion} Quasi-particle dispersions
      $E^{\pm}_{p}$ (top), and occupation numbers $n_a$ and $n_b$
      (middle), in a sample BP state.  This state has gap parameter
      $\Delta \approx 11$ which is the global minimum of the grand
      thermodynamic potential density $\Omega(\Delta)/V$ (bottom) as
      defined in~(\ref{eq:Pressure}).  The maximum at $\Delta \approx
      5.6$ corresponds to an unstable BP state.  These figures
      correspond to the point $(p^F_b,p^F_a) = (11.5,9.2)$ in
      Fig.~\ref{fig:phase}.}
  \end{centering}
\end{figure}
One can now find the \emph{global} minimum by plotting (see
Fig.~\ref{fig:Dispersion})
\begin{equation}
  \label{eq:Pressure}
  \frac{\Omega(\Delta)}{V} = 
  \min_{\braket{\theta|\op{\Delta}|\theta} = \Delta}
  \braket{\theta|\op{\mathcal{H}}-\mu_a\op{n}_a-\mu_b\op{n}_b|\theta},
\end{equation}
where we minimization over all BCS style ansatz $\ket{\theta}$ with
given expectation $\Delta$.  This minimization is equivalent to
comparing \emph{all} solutions of the mean-field gap equation
\begin{equation}
  \label{eq:GapEq}
  \Delta = -\frac{g}{2} \int_R \frac{\d^{3}\vect{q}}{(2\pi)^3}\;
  \frac{\Delta f(q)}{\sqrt{\epsilon_+^2(q)+\Delta^2}}.
\end{equation}
We conclude that, within the mean-field approximation of homogeneous
phases at zero temperature, this model has the phase diagram shown in
Fig.~\ref{fig:phase}.  We plot the properties of a sample BP state in
Fig.~\ref{fig:Dispersion} to illustrate that there are indeed gapless
modes.

To model $\Delta_{p}$ more accurately one might use a function $f(p)$
where the location of the cutoff stays near $p_0$.  This introduces an
inconsistency in the thermodynamics because $f(p)$ is really a
property of the Hamiltonian, while $p_0$ depends on the chemical
potentials $\mu_i$, thus $N \neq -\partial\Omega/\partial\mu$. For
small coupling and high densities, these spurious dependencies become
small and the resulting phase diagram is qualitatively like
Fig.~\ref{fig:phaseP}.

Finally, we address the issue of the instability discussed
in~\cite{Wu:2003} where they claim that the superfluid density is
negative due to a large negative contribution from the diverging
density of states at $E=0$ near the transition to the BP state.  If
one simply computes $\d^2\Omega/\d\Delta^2$, one finds exactly the
same negative contribution indicating that the BP solution under
consideration is an unstable maximum rather than a stable minimum.
The solutions we present here are all global minima, and hence stable.

This raises an interesting point: if the BP/BCS transition were second
order, then the density of states would formally diverge.  Indeed, one
finds $\d^2\Omega(\Delta)/\d\Delta^2 = -\infty$ at certain chemical
potentials.  Near the BP/BCS transitions, $\Omega(\Delta)$ develops a
cusp separating two competing local minima: one is BCS and the other
is BP.  Thus, in the $T=0$ mean-field approximation, the transition
must be \emph{first order}.  At finite temperature, the cusp is
smoothed and we suspect that the transition line ends at a critical
point.  In this way, the $T=0$ BP transition avoids instability.  In
non-extensive systems such as QCD where gapless states may be
stabilized by neutrality constraints, similar instabilities have been
noted~\cite{Huang:2004bg}.  The resolution may be the formation of a
non-homogeneous phase.  This possibility requires further analysis.

Realizations of a stable BP phase require either
non-extensivity, or a finite-range momentum dependent interaction with
a large mass ratio.  The former may occur in high-density
QCD~\cite{Shovkovy:2003uu,Alford:2003fq} where gauge interactions may
stabilize the state.  The latter may occur in a quantum gas of
cold neutral atoms operating near Feshbach resonance with effective
masses tuned by a laser lattice~\cite{Liu:2004mh}, in a system of
trapped ions with dipolar
interactions~\cite{Cirac+Zoller-PhysicsToday:04}, or in
superconductors with overlapping bands~\cite{SMW:1959,Kondo:1963}.
\begin{acknowledgments}
  We would like to thank K.~Rajagopal and C.~Kouvaris for useful
  discussions.  This work is supported in part by funds provided by
  the U.S.  Department of Energy (D.O.E.) under cooperative research
  agreement \#DE-FC02-94ER40818.
\end{acknowledgments}

\end{document}